\documentclass[aps,prl,twocolumn,tightenlines,superscriptaddress,floatfix]{revtex4}
\usepackage{graphicx}
\usepackage{dcolumn}
\usepackage{bm}

\begin{document}

%\preprint{APS/123-QED}

\title{Two-neutron knockout from neutron-deficient $^{34}$Ar,
$^{30}$S, and $^{26}$Si}
\author{K.~Yoneda}
   \altaffiliation[Present address: ]{RIKEN, Hirosawa 2-1, Wako,
     Saitama 351-0198, Japan}
   \affiliation{National Superconducting Cyclotron Laboratory,
     Michigan State University,
     East Lansing, MI 48824}
\author{A.~Obertelli}
   \affiliation{National Superconducting Cyclotron Laboratory,
     Michigan State University,
     East Lansing, MI 48824}
\author{A.~Gade}
   \affiliation{National Superconducting Cyclotron Laboratory,
     Michigan State University,
     East Lansing, MI 48824}
    \affiliation{Department of Physics and Astronomy,
     Michigan State University,
     East Lansing, MI 48824}
\author{D.~Bazin}
   \affiliation{National Superconducting Cyclotron Laboratory,
     Michigan State University,
     East Lansing, MI 48824}
\author{B.\,A.~Brown}
   \affiliation{National Superconducting Cyclotron Laboratory,
     Michigan State University,
     East Lansing, MI 48824}
   \affiliation{Department of Physics and Astronomy,
     Michigan State University,
     East Lansing, MI 48824}
\author{C.\,M.~Campbell}
   \affiliation{National Superconducting Cyclotron Laboratory,
     Michigan State University,
     East Lansing, MI 48824}
   \affiliation{Department of Physics and Astronomy,
     Michigan State University,
     East Lansing, MI 48824}
\author{J.\,M.~Cook}
   \affiliation{National Superconducting Cyclotron Laboratory,
     Michigan State University,
     East Lansing, MI 48824}
   \affiliation{Department of Physics and Astronomy,
     Michigan State University,
     East Lansing, MI 48824}
\author{P.\,D.~Cottle}
   \affiliation{Department of Physics, Florida State University,
     Tallahassee, FL 32306}
\author{A.\,D.~Davies}
   \affiliation{National Superconducting Cyclotron Laboratory,
     Michigan State University,
     East Lansing, MI 48824}
   \affiliation{Department of Physics and Astronomy,
     Michigan State University,
     East Lansing, MI 48824}
\author{D.-C.~Dinca}
   \altaffiliation[Present address: ]{American Science \&
      Engineering, Inc., 829 Middlesex Turnpike, Billerica,
      MA 01821, USA}
   \affiliation{National Superconducting Cyclotron Laboratory,
     Michigan State University,
     East Lansing, MI 48824}
   \affiliation{Department of Physics and Astronomy,
     Michigan State University,
     East Lansing, MI 48824}
\author{T.~Glasmacher}
   \affiliation{National Superconducting Cyclotron Laboratory,
     Michigan State University,
     East Lansing, MI 48824}
   \affiliation{Department of Physics and Astronomy,
     Michigan State University,
     East Lansing, MI 48824}
\author{P.\,G.~Hansen}
   \affiliation{National Superconducting Cyclotron Laboratory,
     Michigan State University,
     East Lansing, MI 48824}
   \affiliation{Department of Physics and Astronomy,
     Michigan State University,
     East Lansing, MI 48824}
\author{T.~Hoagland}
   \affiliation{National Superconducting Cyclotron Laboratory,
     Michigan State University,
     East Lansing, MI 48824}
\author{K.\,W.~Kemper}
   \affiliation{Department of Physics, Florida State University,
     Tallahassee, FL 32306}
\author{J.-L.~Lecouey}
   \altaffiliation[Present address: ]{Laboratoire de Physique
     Corpusculaire, 6 Boulevard du mar\'echal Juin, 14050 Caen Cedex, France}
   \affiliation{National Superconducting Cyclotron Laboratory,
     Michigan State University,
     East Lansing, MI 48824}
\author{W.\,F.~Mueller}
   \affiliation{National Superconducting Cyclotron Laboratory,
     Michigan State University,
     East Lansing, MI 48824}
\author{R.\,R.~Reynolds}
   \affiliation{Department of Physics, Florida State University,
     Tallahassee, FL 32306}
\author{B.\,T.~Roeder}
   \affiliation{Department of Physics, Florida State University,
     Tallahassee, FL 32306}
\author{J.\,R.~Terry}
   \affiliation{National Superconducting Cyclotron Laboratory,
     Michigan State University,
     East Lansing, MI 48824}
   \affiliation{Department of Physics and Astronomy,
     Michigan State University,
     East Lansing, MI 48824}
\author{J.\,A.~Tostevin}
   \affiliation{Department of Physics, School of Electronics and
   Physical Sciences, University of Surrey, Guildford, Surrey GU2 7XH,
  United Kingdom}
\author{H.~Zwahlen}
   \affiliation{National Superconducting Cyclotron Laboratory,
     Michigan State University,
     East Lansing, MI 48824}
   \affiliation{Department of Physics and Astronomy,
     Michigan State University,
     East Lansing, MI 48824}

\date{\today}

\begin{abstract}
Two-neutron knockout reactions from nuclei in the proximity of the
proton dripline have been studied using intermediate-energy beams of
neutron-deficient $^{34}$Ar, $^{30}$S, and $^{26}$Si. The inclusive
cross sections, and also the partial cross sections for the
population of individual bound final states of the $^{32}$Ar,
$^{28}$S and $^{24}$Si knockout residues, have been determined using
the combination of particle and $\gamma$-ray spectroscopy. Similar
to the two-proton knockout mechanism on the neutron-rich side of the
nuclear chart, these two-neutron removal reactions from already
neutron-deficient nuclei are also shown to be consistent with a
direct reaction mechanism.
\end{abstract}


\maketitle

Over the past decade, direct one-nucleon knockout reactions from
fast beams \cite{1nucleon} have been developed into a versatile tool
applicable to structure studies of atomic nuclei beyond the valley
of $\beta$ stability. The technique has been used successfully to
derive single-particle spectroscopic strengths and assign orbital
angular momenta, to probe shell closures, and to study halo nuclei
as well as correlation effects beyond effective-interaction theory
\cite{Nav98,Baz98,Aum00,Gui00,Nav00,Sau00,Mad01,End02,Tos02,Tho03,End03,Gad04a,Gad04b,Ter04,Gad05}.
The associated formalisms, used to calculate single-particle cross
sections and to deduce spectroscopic factors and orbital angular
momenta, employ few-body reaction theory in eikonal approximation
\cite{Tos99} and modern shell-model and Hartree-Fock calculations
\cite{Bro01,Bro98,Bro00}. For several well-studied cases, results
from intermediate-energy knockout reactions have been shown to be
consistent with information obtained from alternative approaches
\cite{brown}.

Two-proton knockout from neutron-rich nuclei has recently been shown
to proceed as a direct reaction \cite{2nucleons}. In that first
reported experiment, two protons were removed from secondary beams
of neutron-rich $^{28}$Mg, $^{30}$Mg, and $^{34}$Si using a thick
$^{9}$Be target. The reaction mechanism, involving sudden,
peripheral collisions, is advantageous for the study of exotic
nuclei since, for a given incoming beam, the residue from two-proton
knockout is even more neutron-rich than that produced by one-proton
removal, but with still manageable cross sections. The method has
also been applied to study the very exotic nuclei $^{42}$Si and
$^{44}$S \cite{fridmann}. A first theoretical description of the
reaction process considered the stripping of two uncorrelated
nucleons \cite{2nucleons}. The correlations between the two removed
nucleons within the initial nucleus have recently been incorporated
via shell-model two-nucleon amplitudes \cite{correl}. These
correlations were found to have a significant influence on the
stripping partial cross sections to different final states,
indicating that two-nucleon knockout reactions offer promise for
accessing nuclear-structure information for very exotic
nuclei and probe two-nucleon correlation effects in asymmetric
regimes \cite{correl}.

\begin{figure}[b]
\includegraphics[width=8.5cm]{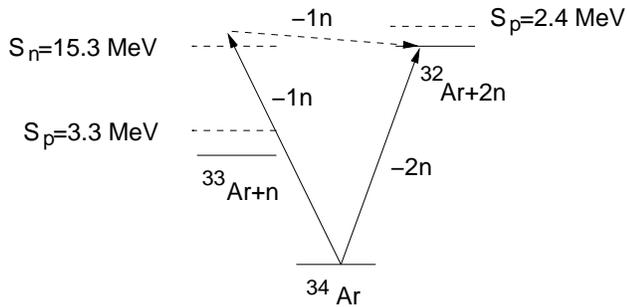}
\caption{Energy thresholds of the direct and two-step processes in
the $^{9}$Be($^{34}$Ar,$^{32}$Ar)X two-neutron removal reaction. The
two-step process would involve neutron evaporation from $^{33}$Ar,
but this channel is strongly suppressed since the proton separation
energy $S_p$ is much lower than the neutron separation energy $S_n$.
Therefore, the two-neutron knockout from $^{34}$Ar to $^{32}$Ar
proceeds predominantly as a direct reaction.} \label{2nko34ar}
\end{figure}

Experiments have, so far, employed only two-proton knockout on the
neutron-rich side of the nuclear chart. The present experiment
provides the first measurements of two-neutron knockout from
proton-rich nuclei. We report first results from the two-neutron
removal reactions $^{9}$Be($^{34}$Ar,$^{32}$Ar)X, $^{9}$Be($
^{30}$S,$^{28}$S)X and $^{9}$Be($^{26}$Si,$^{24}$Si)X in the
proximity of the proton dripline. These nuclei were chosen since (i)
the respective projectiles and knockout residues are expected to be
well described by shell-model calculations within the $sd$ shell,
and (ii) the proton-to-neutron asymmetry in these nuclei strongly
favors a direct two-nucleon removal reaction mechanism. \emph{A
priori}, two-nucleon removal reactions are expected to have both a
direct and a multi-step component. However, the importance of the
latter will depend strongly upon the thresholds for nucleon
emission. For illustration, we consider the $^{34}$Ar projectile
where the dominant two-step, two-neutron removal would involve
neutron evaporation from highly excited states in $^{33}$Ar
(Fig.~\ref{2nko34ar}) formed by single-nucleon removal. However, due
to the neutron-deficiency of $^{33}$Ar, the proton evaporation
channel opens first and neutron evaporation leading to $^{32}$Ar
will be suppressed ($S_p(^{33}$Ar)$=3.3 $~MeV $\ll$
$S_n(^{33}$Ar)$=15.3$~MeV). The same considerations apply to $^{30}$S
and $^{26}$Si. Similar arguments were applied in the work of Bazin
{\it et al.} \cite{2nucleons}.
%, where the nucleon
%thresholds suppress multi-step contributions in two-proton removal
%from neutron-rich nuclei

\begin{figure}
\includegraphics[width=7.8cm]{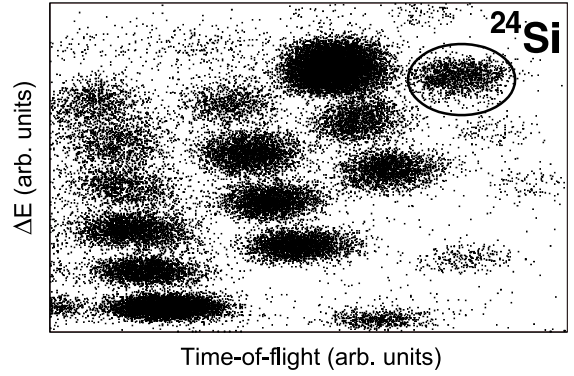}
\caption{Energy-loss vs. time-of-flight identification of the
reaction residues detected in the S800 focal plane for incoming
$^{26}$Si projectiles.
%The energy loss $\Delta$E was measured in
%the ionization chamber of the S800 focal-plane detection system
%and the time-of-flight (ToF) was taken between plastic
%scintillators at the object point of the spectrograph's analysis
%beam line and in the back of the focal plane.
The ToF is corrected for the ion's flight path employing the angle
information obtained from the CRDC detectors in the S800 focal
plane.} \label{ident}
\end{figure}

The projectiles of interest -- $^{34}$Ar, $^{30}$S and $^{26}$Si --
were produced by fragmentation of a 150 MeV/nucleon $^{36}$Ar
primary beam provided by the Coupled Cyclotron Facility of the
National Superconducting Cyclotron Laboratory (NSCL) at Michigan
State University. The $^9$Be production target was located at the
mid-acceptance target position of the A1900 fragment separator
\cite{a1900}.
%The beam purities were 75(2)\%
%($^{34}$Ar), 13(1)\% ($^{30}$S) and 18(1)\% ($^{26}$Si),
%respectively.
The beam impinged on a 376 mg/cm$^2$ $^9$Be secondary target located
at the pivot point of the high-resolution, large-acceptance S800
magnetic spectrograph \cite{s800}. The average mid-target energies
of the beams were 110 MeV/nucleon for $^{34}$Ar, 111 MeV/nucleon for
$^{30}$S and 109 MeV/nucleon for $^{26}$Si.
%The intensities of the beams were
%monitored by scintillators in the beam line upstream of the
%reaction target. The magnetic rigidity of the spectrometer was set
%to transmit the two-neutron knockout residues of interest,
%$^{32}$Ar, $^{28}$S and $^{24}$Si, respectively.
Incoming particles and reaction products were unambiguously
identified on an event-by-event basis. The incident particles were
characterized via their time-of-flight (ToF) taken between two
beam-monitoring scintillators before the target (30 meters apart).
The identification of the reaction residues was performed with the
detection system of the S800 focal plane: the energy loss
($\Delta$E) in the ionization chamber and the ToF measured between
the object point and the focal plane of the spectrograph. The
time-of-flight information was corrected for the flight-path
difference of the various reaction residues using the angle
information provided by the position-sensitive cathode-readout drift
chambers (CRDCs)~\cite{s800} of the S800 focal plane. As an example,
the $\Delta$E-ToF particle identification of $^{24}$Si knockout
residues in the focal plane is shown in Fig.~\ref{ident}.
%The events were selected in coincidence with incoming $^{26}$Si
%beam particles.

The inclusive cross sections for the two-neutron knockout to all
bound final states of $^{32}$Ar, $^{28}$S and $^{24}$Si were derived
from the ratio of detected knockout fragments in the S800 focal
plane relative to the number of incoming projectiles per number
density of the $^{9}$Be target.
%The yields were corrected for dead time,
%detection efficiencies, and for the finite angular and momentum
%acceptance of the spectrograph.
Corrections for the finite acceptance of the spectrograph did not
exceed 15\%. Systematic uncertainties arise from the
particle-identification gate ($\sim$5\%), purity and stability of
the incoming beam ($<$5\%) and  acceptance corrections ($<$10\%).
These uncertainties have been added in quadrature to the statistical
errors.

\begin{figure}
\includegraphics[width=8.0cm]{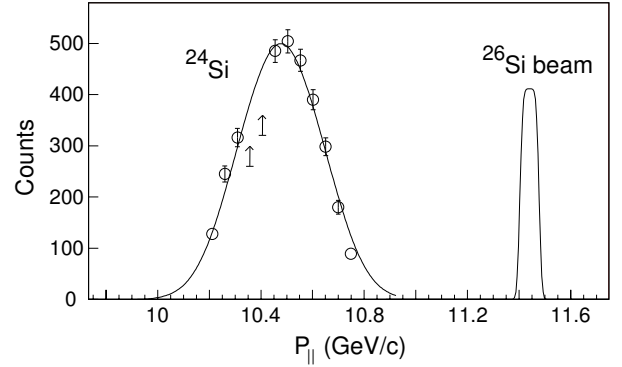}
\caption{Inclusive parallel-momentum distribution for $^{24}$Si
residues from $^{26}$Si incoming particles. The solid line indicates
the theoretical distribution for the removal of two uncorrelated
neutrons, each with $\ell=2$, convoluted with the experimental
momentum spread of the $^{26}$Si beam (shown on the right). The two
data points indicated by lower-limit arrows result from a damaged
electronics chip (confined to a few channels) in one of the
position-sensitive detectors in the S800 focal plane.}
\label{momentum}
\end{figure}

The reaction mechanism has been probed via the parallel-momentum
distributions of the ejectiles measured in the S800 focal plane. The
momentum distributions were reconstructed from the known magnetic
rigidity of the spectrometer and the positions of the particles
measured with the CRDCs. The parallel-momentum distribution of the
$^{24}$Si reaction residues is shown in Fig. \ref{momentum} and
compared to theoretical calculations for the knockout of two
neutrons, each with angular momentum $\ell=2$. The two experimental
points shown as arrows (in the low-momentum half of the
distribution) correspond to a few channels where one of the two
CRDCs was not read out properly during the experiment. The
assumptions made in the theoretical calculation are the same as in
\cite{2nucleons}: the neutrons are considered to be uncorrelated and
so the momentum distribution for the two-neutron knockout is the
convolution of the momentum distributions from removal of each of
the $\ell$=2 neutrons. The theoretical calculation has also been
convoluted with the experimental momentum spread of the incoming
beam ($\sim$ 1 \%). This comparison shows that the theory is in good
agreement with the experimental distribution of the two-nucleon
knockout residues. For one-nucleon removal reactions, the
parallel-momentum distribution of knockout residues is used to
assign the $\ell$-value of the removed nucleon. In two-nucleon
knockout reactions, the parallel-momentum distributions for
different $\ell$-values can be too similar to provide a clear
assignment of the orbital angular momentum.

%\begin{figure}
%\includegraphics[width=8.5cm]{setup_n.eps}
%\caption{Schematic of the experimental setup with SeGA in front of
%the S800 spectrometer.}
%\label{setup}
%\end{figure}

\begin{figure}
\includegraphics[width=8.3cm]{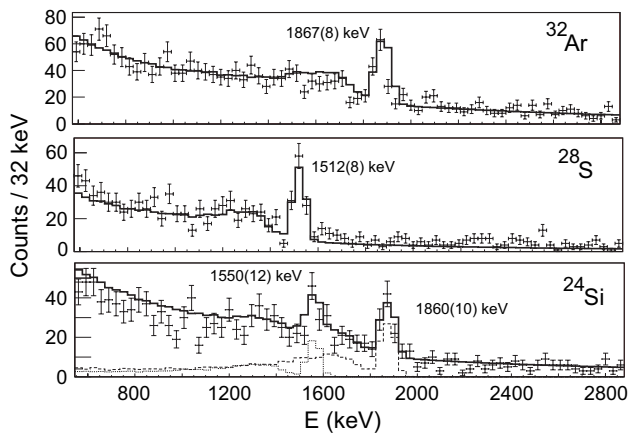}
\caption{$\gamma$-ray spectra reconstructed event-by-event into the
rest frame of the emitting nucleus, in coincidence with $^{32}$Ar,
$^{28}$S and $^{24}$Si residues. Statistical error bars are
indicated. The solid line is the result of a GEANT \cite{geant}
simulation. For $^{24}$Si, the simulated response for the two
photopeaks is shown separately in dashed an dotted lines.}
\label{fig1}
\end{figure}

Cross sections to individual excited states were measured using
particle-$\gamma$  coincidences. The target was surrounded by the
SeGA~\cite{sega}, an array of seventeen 32-fold segmented
high-purity Germanium detectors arranged in two rings as described
in~\cite{Obe06}.
%at a distance of 20.8 cm from the center of the
%target. Seven detectors were positioned at forward angles in a
%37$^{\circ}$ ring and ten detectors were located at 90$^{\circ}$
%relative to the beam axis. The high degree of segmentation allowed
%event-by-event Doppler-reconstruction of $\gamma$ rays emitted by
%nuclei in-flight. The photopeak efficiency of SeGA was 2.5\% for a 1.3
%MeV transition emitted at a velocity of $v/c=0.35$.
%The germanium
%detectors were calibrated in energy with $^{56}$Co, $^{152}$Eu and
%$^{226}$Ra sources. The efficiency of the setup was simulated with the
%GEANT code~\cite{geant} to include the Lorentz boost and verified
%against calibrated-source measurements.
The event-by-event Doppler-corrected $\gamma$-ray spectra for the
$^{32}$Ar, $^{28}$S and $^{24}$Si residues are presented in
Fig.~\ref{fig1}. The Doppler correction has been performed by taking
into account the average velocity of the projectile at the time of
the $\gamma$-ray emission.
%The mean lifetimes of excited
%states can lead to the $\gamma$ rays being emitted mid-target, in
%the second half of the target or even outside the target, giving
%rise to different velocities at the time of the $\gamma$-ray
%emission for different transitions. Each observed $\gamma$-ray
%peak has been confirmed to be aligned in energy for the two
%angle groups, ensuring an optimum result for the Doppler
%reconstruction. The large number of fragmentation products
%detected in the focal plane offered several well-known transitions
%for which the target position relative to the center of SeGA could
%be accurately determined, thus leaving the velocity $v/c$ as main
%parameter to be optimized for the Doppler reconstruction.
Although an anisotropic angular distribution is expected, due to
alignment effects in the knockout reaction, we assume that this can
be neglected in the evaluation of the intensities. The smallness of
this correction is tied to the beam energy and to the particular
choice of laboratory angles for the $\gamma$-ray detectors
(37$^{\circ}$ and 90$^{\circ}$) in this experiment; see the example
worked out in Fig. 12 of Ref.~\cite{1nucleon} for one-nucleon
knockout.

In the case of $^{32}$Ar ($S_p=2.4$~MeV), only one $\gamma$-ray
transition is observed at 1867(8) keV corresponding to the decay of
the first 2$^{+}$ excited state. The measured energy is slightly
different from the energy of 1824(12) keV reported in \cite{cottle}
using a scintillator array. We obtain an inclusive cross section of
$\sigma=0.48(6)$ mb for the two-neutron knockout and a 0.07(4)~mb
cross section to the first 2$^+$ excited state. Assuming no other
bound excited states, the ground state is fed directly with a cross
section of 0.41(7)~mb.
%The
%quoted uncertainty of the ground-state cross section is the square
%root of the quadratic sum of the uncertainties of the inclusive
%cross section and the cross section for the excited state.

\begin{figure}
\includegraphics[width=8.5cm]{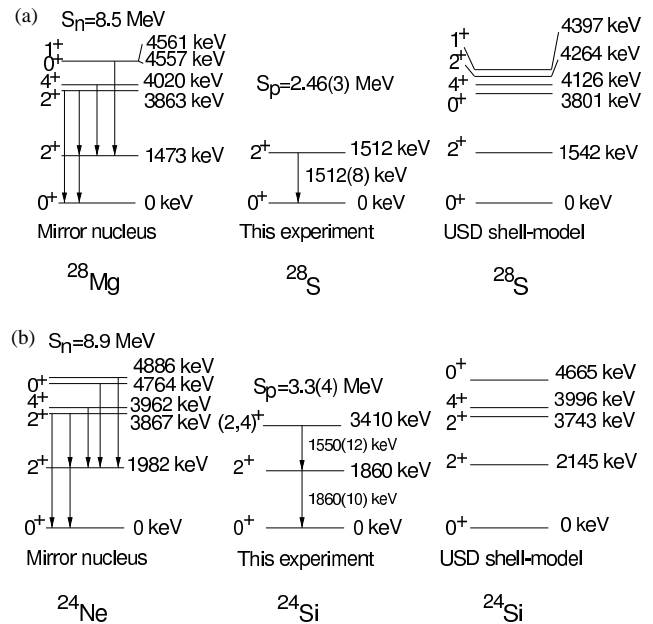}
\caption{Level schemes of $^{28}$S (a) and $^{24}$Si (b). The
experimental results (center) are compared to shell-model
predictions and to the level scheme of the respective mirror
nucleus.} \label{schemes}
\end{figure}

For $^{28}$S ($S_p=2.46(3)$~MeV), a new transition has been observed
at 1512(8) keV. This transition is assigned to the decay of the
first 2$^+$ state of $^{28}$S, based on shell-model calculations and
comparison to the mirror nucleus. A shell-model calculation with the
Oxbash code \cite{oxbash} and the USD interaction \cite{usd}
predicts the first 2$^+$ state at 1543 keV excitation energy, while
that of the mirror nucleus, $^{28}$Mg, is at 1473~keV (see Fig.
\ref{schemes}(a)). The inclusive cross section for the production of
$^{28}$S from $^{30}$S is $\sigma= 0.73(8) $~mb. The 2$^+$ excited
state is populated with a 0.34(8) mb cross section leaving a
0.39(8)~mb cross section for the knockout to the ground state.

Of the three nuclei studied, $^{24}$Si is the most strongly bound
with a proton separation energy of $S_p= 3.3(4)$ MeV. Two
$\gamma$-ray transitions are observed, at 1550(12) keV and 1860(10)
keV, the latter being about twice as intense as the former. The
observed transitions correspond to the decay of the two previously
reported excited states of $^{24}$Si at 3441(10) keV and 1879(11)
keV \cite{schatz}, respectively. The statistics are too low to allow
a $\gamma$-$\gamma$ coincidence study of these two transitions. The
mirror nucleus $^{24}$Ne exhibits a vibrator-like excitation scheme
with a first 2$^+$ state lying at 1981.6(4) keV and a (2$^+$,4$^+$)
doublet at 3867(8) keV and 3962(18) keV~\cite{howard}, respectively.
Each state of the doublet decays with an almost 100\% branch to the
first excited state. A comparison with $^{24}$Ne suggests that the
two lines observed for $^{24}$Si in the present work are in
coincidence with the first 2$^+$ state at 1860(10) keV and a
(2,4)$^+$ level at 3410(16) keV. The resulting level scheme is shown
in Fig. \ref{schemes} (for a discussion on mirror asymmetry in
these nuclei see~\cite{Her95}). Within the energy resolution of
the present setup, there is no indication of a doublet at around
1550~keV.   

%\begin{figure}
%\includegraphics[width=8.5cm]{fig6}
%\caption{Level scheme of $^{24}$Si. Same as caption for
%Fig.~\ref{scheme28s}. } \label{fig2}
%\end{figure}

Our experimental results are summarized in Table \ref{results}. The
measured inclusive and partial cross sections are given. For
comparison, we also include the inclusive cross sections for the
two-proton removal reactions reported in \cite{2nucleons}. The
different reactions show cross sections of the order of 1 mb for
two-proton (two-neutron) knockout from neutron-rich
(neutron-deficient) nuclei in the $sd$-shell.

\begin{table}
\begin{center}
\vspace{0.5cm} \caption{Spin, excitation energy of the final states,
and experimental cross sections for the two-neutron knockout
reactions $^{9}$Be($^{34}$Ar,$^{32}$Ar)X, $^9$Be($
^{30}$S,$^{28}$S)X and $^9$Be($^{26}$Si,$^{24}$Si)X. The results are
compared to the theoretical cross sections which are broken down
according to contributions from stripping, stripping-diffraction and
diffraction. For comparison, the inclusive cross sections from the
two-proton knockout reactions $^9$Be($^{34}$Si,$^{32}$Mg)X,
$^9$Be($^{30}$Mg,$^{28}$Ne)X and $^9$Be($^{28}$Mg,$^{26}$Ne)X
\protect\cite{2nucleons} are also presented.}
\begin{ruledtabular}
\begin{tabular}{cccc|c|ccc}
Proj.&$J^{\pi}_{f}$&$E^{\star}_{\rm exp}$&$\sigma_{\rm exp}$&$\sigma_{\rm th}$&str&str-diff&diff\\
& &(keV)&(mb)&(mb)&(\%)&(\%)&(\%)\\
\hline
$^{34}$Ar&0$^{+}_{gs}$&0&0.41(7)&0.71&54&39&7\\
&2$^{+}$&1867(8)&0.07(4)&0.35&51&41&8\\
&&inclusive&0.48(6)&1.06&  &\\
\hline
$^{30}$S&0$^{+}_{gs}$&0&0.39(8)&0.84&55&38&7\\
&2$^{+}$&1512(8)&0.34(8)&0.69&59&36&5\\
&&inclusive &0.73(8)&1.54& & &\\
\hline
$^{26}$Si&0$^{+}_{gs}$&0&0.71(9)&1.30&55&39&7\\
&2$_1^{+}$&1860(10)&0.15(4)&0.30&60&35&5\\
&($4_1$,2$_2$)$^{+}$&3410(16)&0.14(4)&0.30&61&34&5\\
&&inclusive&1.01(10)&1.90& & &\\
\hline
\multicolumn{8}{c}{Inclusive two-proton removal cross sections from \protect\cite{2nucleons}}\\
\hline
$^{34}$Si&&inclusive&\multicolumn{5}{l}{0.76(10)}\\
$^{30}$Mg&&inclusive&\multicolumn{5}{l}{0.49(5)}\\
$^{28}$Mg&&inclusive&\multicolumn{5}{l}{1.50(10)}\\
\end{tabular}
\end{ruledtabular}
\label{results}
\end{center}
\end{table}

Our theoretical calculations of the two-neutron-removal cross
sections follow the formalism and notation of Ref. \cite{correl}, 
which developed a full treatment of the two-nucleon stripping 
(absorption) cross section, $\sigma_{str}$. Here, we also include 
a full calculation of contributons to the cross sections from events 
where only one of the nucleons is stripped (absorbed) and the second 
is removed by an elastic collision (diffraction of the nucleons or 
residue) with the target. This is denoted $\sigma_{\rm str-diff}$. We 
only estimate the small cross section, $\sigma_{\rm diff}$, from events 
in which both of the tightly-bound nucleons are removed by elastic 
dissociation.

The two knocked-out nucleons are assumed to be removed from a set of
active and partially occupied single-particle orbitals $\phi_{j}$, 
with spherical quantum numbers $n(\ell j)m$. The assumed two-nucleon
overlap function, of the initial state $J$ with each residue final
state $f$, is denoted by $\Psi_{J M}^{(f)}$ \cite{correl}. In the
eikonal model of the direct reaction dynamics, the ($A$-2)-body
residue (or core) is a spectator and is assumed to interact at most 
elastically with the target. It thus enters the formalism through a 
residue-target elastic transmission probability $|{\cal S}_c|^2$.
Events in which a nucleon is absorbed are described by the
nucleon-target absorption probabilities $[1-|{\cal S}_i|^2]$.
The ${\cal S}$ and $\phi_{j}$ are calculated as discussed in Ref.\ 
\cite{Gad04b}, being constrained by Hartree-Fock systematics.

Contributions to the two-nucleon knockout cross section from the
diffractive removal of one nucleon, say 1, and absorption of the 
second, 2, are included in the expression
\begin{equation}
\sigma_{1}=\frac{1}{2J+1}\sum_{M} \int d\vec{b} \langle \Psi_{J M}^{(f)}| 
|{\cal S}_c|^2 |{\cal S}_1 |^2 [1-|{\cal
S}_2|^2]| \Psi_{J M}^{(f)}\rangle , \label{sum1}
\end{equation}
and similarly for diffraction of nucleon 2. However, as currently 
stated, Eq.\ (\ref{sum1}) includes events in which nucleon 1 remains 
bound to the residue. These single-nucleon removal events, populating 
bound ($A$-1)-body residues, must be removed by projecting off
the nucleon-residue bound states, by replacing in Eq.\ (\ref{sum1})
\begin{eqnarray}
|{\cal S}_1|^2 \rightarrow {\cal S}_1^* \LARGE[1-\sum_{j''m''}
|\phi_{j''}^{m''} ) ( \phi_{j''}^{m''} |\, \LARGE] {\cal S}_1~.
\label{complete}
\end{eqnarray}
Here the sum is over the bound eigenstates $n(\ell'' j'')m''$ of 
nucleon 1 and the core and we use the bra-kets $(..|$ and $|..)$ to 
indicate states and integration over this nucleon's coordinates. 
We include all the active single particle orbitals in this sum. 
After having made this replacement then $\sigma_{\rm str-diff}$=$
\sigma_{1}$+$\sigma_2$.

We only estimate the cross section from removal of both
tightly-bound nucleons by elastic dissociation. We make use of
the reduction in the removal cross section when one nucleon is 
dissociated rather than stripped, i.e. $\sigma_{\rm i}/\sigma_{\rm str}$ 
as calculated above. We thus estimate the two-nucleon diffractive
cross section to be $\sigma_{\rm diff}$$\approx$$[\sigma_{\rm
    i}/\sigma_{\rm str}]^2
\sigma_{\rm str}$. Since, see Table \ref{results}, $\sigma_{\rm
  i}/\sigma_{\rm str} 
\approx 0.35 - 0.4$, we estimate that $\sigma_{\rm diff}$ makes a 
contribution of only $6-8\%$ to the two-nucleon removal partial 
cross sections. Finally, the theoretical cross sections are the 
sum $\sigma_{\rm th} = \sigma_{\rm str}+ \sigma_{\rm str-diff}+
\sigma_{\rm diff}$.
For all of the three systems studied these $\sigma_{\rm th}$ over-predict 
the measured cross sections by about a factor of two, requiring an
empirical suppression of the two-neutron shell model strengths. This 
reduction - somewhat analogous to the suppressions observed in 
single-nucleon removal reactions - will be discussed elsewhere
\cite{tostevin}.

\begin{figure}
\includegraphics[width=8.5cm]{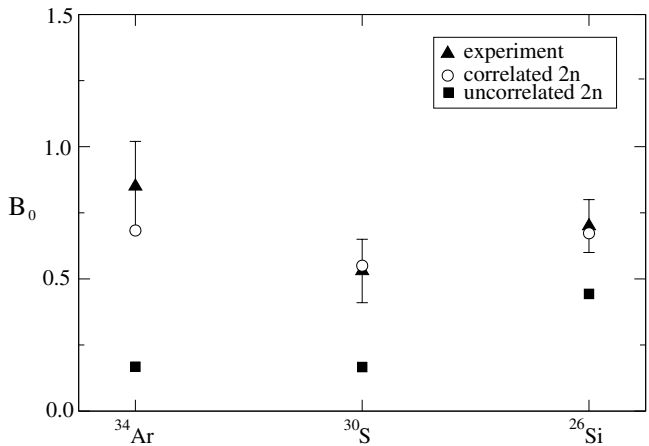}
\caption{Ground state branching ratios $B_{0}=\sigma(0^+)/\sigma_{
\rm incl}$ in the two-neutron removal reactions from $^{26}$Si,
$^{30}$S and $^{34}$Ar from (i) the experiment, (ii) calculated
assuming the removal of two uncorrelated neutrons, and (iii) when
including correlations via the many-body shell model wave functions
\protect\cite{correl,tostevin}.} \label{fig3}
\end{figure}

The observed differences in the cross sections to individual final 
states suggests (i) significant sensitivity of the reaction to the 
single-particle structures of the projectiles and residues, and
thus (ii) that the direct two-nucleon removal mechanism provides 
opportunities to probe aspects of nuclear structure far from stability.
Figure \ref{fig3} shows the ground-state transition branching ratios,
$B_{0}=\sigma(0^+)/\sigma_{\rm incl}$, of the three two-neutron removal
reactions. The measured $B_{0}$ are compared to two model calculations. 
The first assumes the knockout of two completely uncorrelated neutrons 
from pure configurations $\nu [d_{5/2}]^{4}$, $B_0$=4/9, $\nu [d_{5/2}
]^{6}$, $B_0$=1/6, and $\nu [d_{5/2}]^{6}[s_{1/2}]^{2}$, $B_0$$\approx$1/6, 
for $^{26}$Si, $^{30}$S and $^{34}$Ar, respectively (see e.g. 
Section III of \cite{correl}). The second calculation includes fully
the pairing correlations between the neutrons as given by the many-body, 
$sd$-shell model wave functions \cite{correl,tostevin}. The experimental 
branching ratios are in agreement with the model for removal of two 
correlated neutrons while the uncorrelated neutrons assumption fails 
to reproduce the data. Specifically, these direct two-neutron knockout 
reactions show sensitivity to, and insight into the pair-correlations 
of the two neutrons, leading to an enhanced $0^+$ cross section. Unlike 
one-nucleon knockout reactions, here the two-particle shell-model 
components contribute coherently for a given total angular momentum, 
resulting in interference effects. This strong interplay between nuclear 
structure and the reaction dynamics is evident in the results in 
Table~\ref{results} and Fig.~\ref{fig3}.

In summary, two-neutron knockout reaction measurements have been
performed on the already neutron-deficient nuclei $^{34}$Ar,
$^{30}$S, and $^{26}$Si, near the proton dripline. Particle-$\gamma$
coincidences allowed the measurement of the partial cross sections
to individual bound final states. Level schemes of the residues are
proposed using comparisons with the mirror nuclei and with USD
shell-model calculations. The inclusive two-neutron removal cross
sections from these proton-rich nuclei with $N=12,14,16$ were found
to be of the order of 1~mb; similar to those observed in two-proton
knockout experiments from the neutron-rich $sd$-shell nuclei with
$Z=12,14$ \cite{2nucleons}. Based on the energetics of the reaction
and the observed cross sections, the two-neutron removal processes
discussed here appear to proceed by the direct reaction mechanism.

\begin{acknowledgments}
This work was supported by the U.S. National Science Foundation
under Grants No. PHY-0110253 and No. PHY-0244453 and by the United
Kingdom Engineering and Physical Sciences Research Council (EPSRC)
Grant No. EP/D003628.
\end{acknowledgments}

\end{document}